\documentclass[preprint]{revtex4}

\usepackage{graphicx}% Include figure files
\usepackage{dcolumn}% Align table columns on decimal point
\usepackage{bm}% bold math
\usepackage{epsfig}
\usepackage{footnote}
\usepackage{times}

\begin{document}

\title{Quantum key distribution at telecom wavelengths with noise-free detectors}

\author{Danna Rosenberg}\thanks{Present address: Los Alamos
National Laboratory, Los Alamos, NM}
\author{Sae Woo Nam}%
\affiliation{National Institute of Standards and Technology, $325$ Broadway, Boulder, CO 80305}%

\author{Philip A. Hiskett}%
\author{Charles G. Peterson}%
\author{Richard J. Hughes}
\author{Jane E. Nordholt}%
\affiliation{Los Alamos National Laboratory, Los Alamos, NM}

\author{Adriana E. Lita}%
\author{Aaron J. Miller}\thanks{Present address: Physics Department, Albion College, Albion, MI}%
\affiliation{National Institute of Standards and Technology , $325$ Broadway, Boulder, CO 80305}%

\date{\today}% It is always \today, today,

\begin{abstract}
The length of a secure link over which a quantum key can be
distributed depends on the efficiency and dark-count rate of the
detectors used at the receiver.  We report on the first
demonstration of quantum key distribution using transition-edge
sensors with high efficiency and negligible dark-count rates.
Using two methods of synchronization, a bright optical pulse
scheme and an electrical signal scheme, we have successfully
distributed key material at $1,550$~nm over $50$~km of optical
fiber. We discuss how use of these detectors in a quantum key
distribution system can result in dramatic increases in range and
performance.
\end{abstract}


\maketitle

When properly implemented, quantum key distribution (QKD) provides
a means of secure communication where privacy is guaranteed by the
laws of physics rather than by computational complexity
\cite{roadmap}. The sender of information (Alice) in a QKD system
encodes information in the state of a single photon before sending
it to the receiver (Bob). The action of an eavesdropper (Eve)
alters the quantum state of the photon, making her presence known
to Alice and Bob. "Prepare-and-measure" QKD protocols such as BB84
\cite{BB84} that utilize a single-photon-on-demand source can be
unconditionally secure \cite{Mayers01}. Much research is being
performed in the field of single-photon sources (see
\cite{singlephotonsources} for a comprehensive review), but at
present such sources are not readily available. Instead, QKD
systems often use a heavily attenuated laser pulse to approximate
a single photon source. However, the presence of multi-photon
signals provides Eve with new opportunities; she could remove a
photon from each multi-photon signal and measure it without being
detected \cite{Brassard00}.

The number of photons per pulse from a heavily attenuated laser
source obeys a Poisson distribution, so the probability of a
multi-photon event in a pulse with mean photon number $\mu < 1$ is
approximately $\mu^2/2$. It has been established that BB84 QKD
with weak laser pulses can be unconditionally secure
\cite{Inamori01} but requires $\mu<\eta$, where $\eta$ is the
channel transmittance \cite{Brassard00}.  However, detector dark
counts place a practical lower limit on $\mu$ (or, similarly, the
length of a secure link) because eventually Bob is more likely to
get a dark count than a photon with which Alice has encoded
information. Detectors with high efficiency, low dark-count rate,
and short recovery times can both improve the security of a QKD
link and allow a key to be transmitted over a longer length of
optical fiber. Such detectors would have similar advantages for
new, "decoy state" protocols \cite{Hwang03}, which permit secure
operation at mean photon numbers $\mu \sim 1$.

Transition-edge sensors (TESs) are sensitive microcalorimeters
that detect photons by measuring the temperature rise of a small
superconducting sample due to the absorption of individual photons
\cite{Irwin95,Cabrera98}. Unlike single-photon-sensitive avalanche
photo-diodes (APDs) that are typically used in fiber QKD systems
\cite{Hiskett00}, they can operate at the telecommunications
wavelengths of $1310$~nm and $1550$~nm with high efficiency and
negligible dark-count rates. The TESs used in this work are
sensitive to single near-infrared photons and consist of $20$-nm
thick $25~\mu$m square tungsten detectors each embedded in a stack
of optical elements designed to maximize the detection efficiency
at $1550$~nm. The recovery time after a detection event is
$4~\mu$s.  Note that this time is significantly shorter than
typical APD afterpulse blocking times and could result in higher
secret bit rates in QKD systems that have been demonstrated using
APDs. These sensors can detect $1550$~nm photons with an
efficiency of $89$\% \cite{pra05}, but they are also sensitive to
the presence of room-temperature blackbody radiation that is
transmitted through the optical fiber. The blackbody radiation can
result in a background of up to several hundred counts per second,
but these photons are broadly distributed in energy and can be
filtered out. We filtered the blackbody counts by inserting a
segment of bent single-mode fiber at a mid-temperature stage of
the refrigeration system. The bent fiber, which preferentially
sheds the longer wavelength photons, lowered the detection
efficiency at $1550$~nm from $89$\% to $65$\%, but it also reduced
the rate of blackbody photons from $400$~Hz to approximately
$27$~Hz.

The phase-encoding fiber-based QKD system used here is similar to
other one-way implementations in which Alice and Bob share two
halves of a time-multiplexed Mach-Zender interferometer
\cite{Townsend93,Hughes00}; it will be described in greater detail
elsewhere \cite{Hiskett_tbp}. In these schemes quantum information
is encoded on a photon's optical phase, but this can only be
accomplished with 50\% efficiency owing to the time-multiplexing
of the interferometer onto a single fiber, decreasing the protocol
efficiency. Photons that convey no quantum information arrive at
Bob's detector several nanoseconds earlier or later than the
photons of interest, appearing as side lobes in the arrival time
histogram. Because the TESs do not have sufficient timing
resolution to determine the arrival time of a photon to within
several nanoseconds, removal of these photons was necessary for
operation of the system. We used a technique incorporating an
optical switch at the input to Bob's half of the interferometer to
prevent the photons from taking non-interference paths
\cite{Hiskett_tbp}. Alice's and Bob's phase modulators were placed
outside of each interferometer for increased stability, and the
fiber link consisted of $50$~km of dark single-mode low-dispersion fiber.

We implemented the BB84 protocol \cite{BB84}, in which Alice
randomly encodes a $0$ or a $1$ in either of two conjugate bases
by setting her phase modulator to one of four values. Similarly,
Bob sets his phase modulator to one of two possible values to
select his measurement basis.  After Alice transmits a sequence of
quantum signals, Alice and Bob share their choice of bases (but
not the bit values) over a public channel and discard bits for
which their bases did not match, creating the sifted key. Error
correction \cite{Brassard94} and privacy amplification
\cite{Bennett95} are then performed to obtain the secret key. The
protocol generally requires a detector at each of Bob's two
outputs, one for each bit value. However, rather than use two
detectors with different detection efficiencies and dark count
rates, we chose to time-multiplex the signals from the two paths
onto a single detector \cite{Townsend93} by adding a $320$~ns
optical delay to one of the paths and recombine the photons at a
polarizing beam combiner. The arrival time of the photon within
the timing window set by the clock rate provides the information
about whether the bit was a zero or a one.

The system was clocked at $1$~MHz, and synchronization between
Alice and Bob was performed using two different methods.  We first
discuss a method involving a bright $1310$-nm pulse that precedes
the $1550$-nm pulse, and the limitations of this method. Then, we
show data using electrical synchronization between Alice and Bob.

\begin{figure}[b]
\includegraphics[width=2.5in]{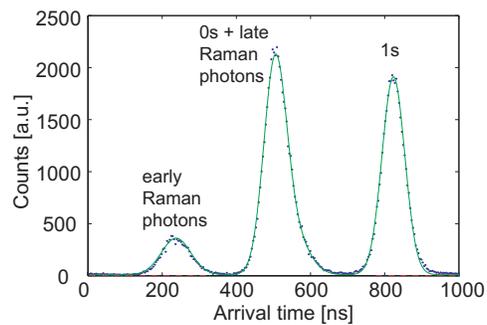}
\caption{(Color online). Histogram of arrival times using a bright
synchronization pulse. The dots are the measured arrival time and
the solid line is the fit to the data discussed in the text.}
 \label{fig:timehist}
\end{figure}

Synchronization using a bright pulse reduces errors from timing
jitter due to slight changes in the length or optical properties
of the fiber link, because both the $1550$~nm light that encodes
the quantum information and the synchronization pulse travel
through the same link.  For this technique to be successful with
TESs, it is important that none of the photons from the
synchronization pulse reach the detectors. Five wavelength
division multiplexers (WDMs) each with 27~dB extinction were
placed at the input to Bob's interferometer to remove the
$1310$~nm light. However, the $1310$~nm photons also create
$1550$~nm photons through Raman processes in the fiber
\cite{Toliver04,Magiq05}. The WDMs do not aid in filtering these
photons, but a $1$~nm bandwidth filter at the input to Bob's
detectors sufficiently reduced the rate of these photons so that
the system was usable. Insertion of this filter resulted in an
extra $3.2$~dB of loss at $1550$~nm. Even with the filter, counts
from the Stokes Raman scattered photons are still evident in the
arrival time histogram in Bob's detector, shown in Fig.
\ref{fig:timehist}. The two main peaks are from photons from the
$1550$~nm laser that carry the quantum information.  The two
smaller peaks (one of which overlaps with the earlier of the main
peaks) are due to the Raman processes discussed above. The ratio
of the areas and the spacing in time of the two larger peaks are
the same as for the two smaller peaks. Therefore, we were able to
perform a fit to the center peak by shifting and scaling the
Gaussian fits to the early clock photons and the 1s bits. We
obtained the best fits with a delay of $319.5$~ns and a ratio of
$1.04$.

\begin{figure}[tb]
\includegraphics[width=2.5in]{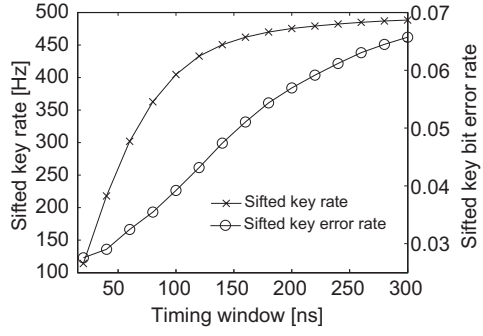}
\caption{Dependence of sifted bit error rate and sifted key rate
on the width of the timing window for data using optical
synchronization.} \label{fig:timewindow}
\end{figure}

The $72$~ns FWHM arrival time resolution seen in Fig.
\ref{fig:timehist} is due to the intrinsic time constant of the
detector. We can reduce the bit error rate by accepting events
only within a narrow time window, at the cost of reducing the
overall key rate. Figure \ref{fig:timewindow} shows the dependence
of these two quantities on timing window width.  The error rate
was obtained by comparing all of Alice and Bob's sifted bits.  For
the sifted key rate and sifted key bit error rate data shown in
this paper, we chose timing windows equal to the FWHM ($72$~ns)
obtained from the timing histogram. Because the ultimate
performance of a QKD system depends on the number of secret bits
transmitted per unit time, optimal timing windows could be chosen
to maximize the secret bit rate. The detection rate of Raman
scattered photons from the clock pulse within the $72$~ns windows
was $30.3$~Hz.

\begin{figure}[tb]
\includegraphics[width=2.75in]{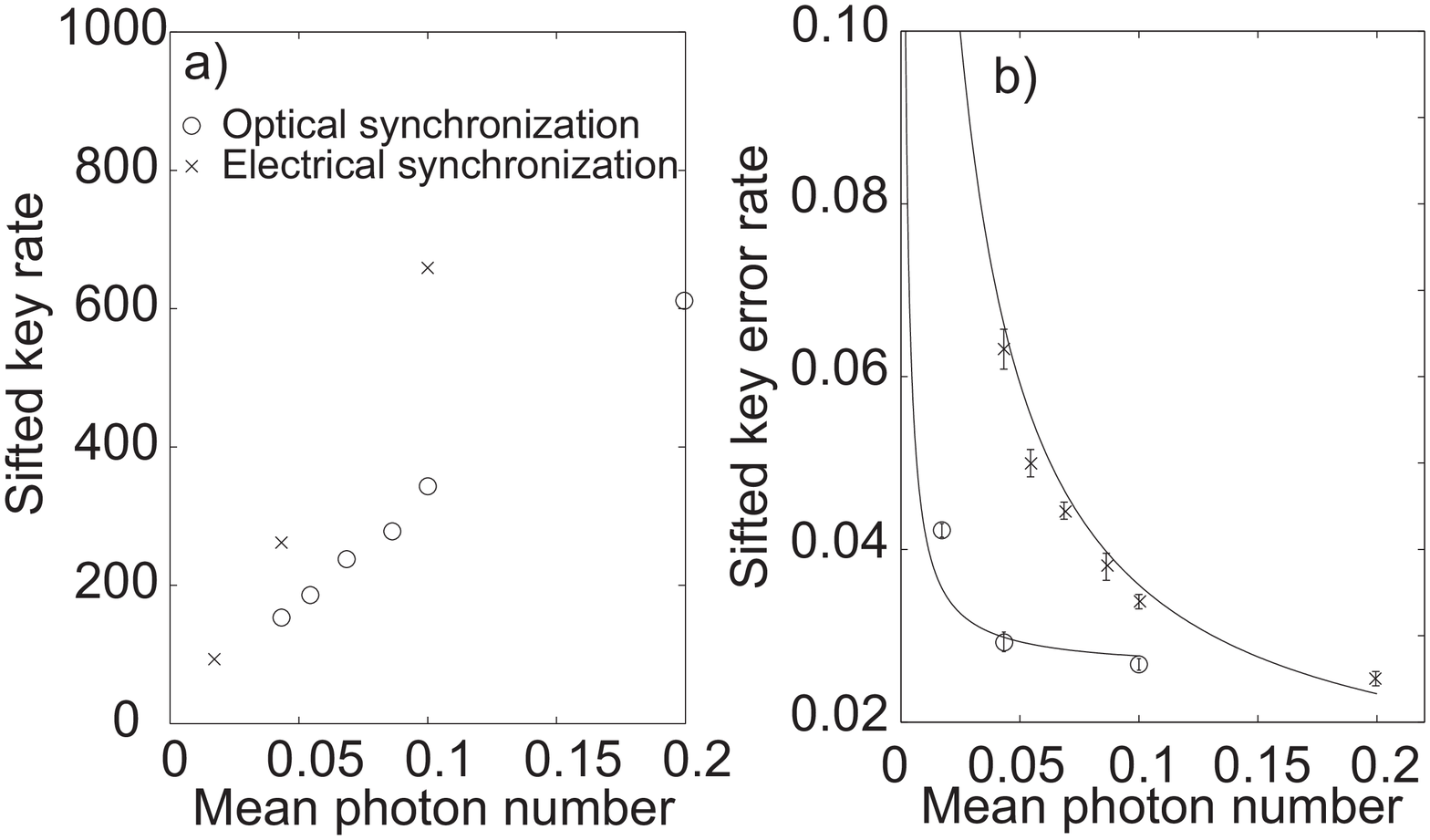}
\caption{Sifted key rate (a) and error rate (b) as a function of
mean photon number for $50$~km of fiber.} \label{fig:elec_vs_opt}
\end{figure}

We also performed experiments using electrical synchronization
with a Rubidium atomic clock.  The experimental setup for these
experiments was similar to that described above, but the $1310$~nm
laser and filter at Bob's input were removed. Figure
\ref{fig:elec_vs_opt} shows the sifted key rate and sifted key bit
error rate for the experiments using optical and electrical
synchronization.  The increase in sifted key rate by a factor of
two for the data using electrical synchronization was due to the
removal of the filter at the detector input.  The sifted key bit
error rate with electrical synchronization is reduced due to two
effects. First, the Stokes photons from the synchronization pulse
that had contributed to the error rate because they overlapped
with the 0~s peak in Fig. \ref{fig:timehist} were removed. Second,
the removal of the filter reduced the losses in Bob's section,
increasing the number of error-free bits detected by Bob,
resulting in a factor of $2.1$ higher sifted key rate. The solid
lines in Fig. \ref{fig:elec_vs_opt} are fits to the experimental
data, using the measured values for the number of Stokes and
background photons present within the $72$~ns timing windows.
Following \cite{Lutkenhaus00} and \cite{Shields04}, we can
calculate the mean photon number $\mu$ at which the sifted key
error rate is $11$\%, the upper limit for the creation of secret
key.  The minimum value of $\mu$ we obtain for a fiber link of
$50$~km can be used to calculate a maximum distance over which we
can transmit secret key using a mean photon number of $\mu=0.1$.
The minimum mean photon numbers (maximum transmission distances)
for the data with optical and electrical synchronization are $2.19
\times 10^{-2}$ ($83$~km) and $1.68 \times 10^{-3}$ ($138$~km),
respectively.

As discussed earlier, the background counts in the detector are
dominated by the leakage of blackbody radiation through the
optical fiber, and these photons can be filtered, resulting in far
lower background counts.  This is fundamentally different from the
case with other detectors, where the dark-count rate may be low,
but it is intrinsic to the detector and cannot be lowered using
filters.  In this paper, we have used a bent-fiber filter, which
is a crude method of filtering these photons.  However, we can now
quantify the advantages to using a better filter. For example,
consider a detector with $89$\% efficiency and $400$~Hz background
count rate \cite{pra05} used with a filter with $10$~nm bandwidth
centered around $1550$~nm, $3$~dB insertion loss in the passband,
and $40$~dB out-of-band rejection. The calculated rate of photons
from the blackbody distribution between $1545$ and $1555$~nm that
couple into single-mode fiber is $0.03$~Hz.  It is difficult to
calculate the rate of the longer wavelength photons because these
photons are so sensitive to bends in the optical fiber, but we can
use the measured value of $400$~Hz of the background rate as an
overestimate.  The total background count would therefore be
$0.053$~Hz, taking into account the finite detector efficiency and
the $3$~dB filter loss at $1550$~nm.  If we assume that the error
due to improper phase modulation and interferometer visibility is
$1$~\%, we find that we can distribute key securely over $271$~km
at a mean photon number of $0.1$.

In summary, we have demonstrated the use of transition-edge
sensors, a novel type of detector with high efficiency and
virtually no dark counts at telecommunication wavelengths, in a
quantum key distribution system.  We have performed experiments
with synchronization by both a bright optical pulse and an
electrical signal, and we have shown how incorporation of these
detectors in a system with improved filtering can result in
impressive increases in the secret bit rate and the length over
which a secure key can be transmitted.

The authors thank ARDA and DARPA for financial support, Alan
 Migdall for the loan of an optical switch, Mark Peters for
 supplying cryogens, and Joe Dempsey and Corning Inc. for
 supplying the optical fiber. D. R. is supported by the DCI
 postdoctoral program.

\end{document}